\documentclass[preprint2]{aastex}
\usepackage{float}
\newcommand{\myemail}{henrikj@astro.lu.se}

\shorttitle{Fluorine in the Solar neighborhood}
\shortauthors{J\"onsson et al.}

\begin{document}

\title{Fluorine in the Solar neighborhood:\\No evidence for the neutrino process}

\author{H. J\"onsson\altaffilmark{1}}
\affil{Lund Observatory, Department of Astronomy and Theoretical Physics, Lund University, Box 43, SE-22100 Lund, Sweden}
\affil{Instituto de Astrof\'{i}sica de Canarias (IAC), E-38205 La Laguna, Tenerife, Spain}
\affil{Universidad de La Laguna, Dpto. Astrof\'{i}sica, E-38206 La Laguna, Tenerife, Spain}
\email{\myemail}
\author{N. Ryde\altaffilmark{1}}
\affil{Lund Observatory, Department of Astronomy and Theoretical Physics, Lund University, Box 43, SE-22100 Lund, Sweden}
\author{E. Spitoni and F. Matteucci}
\affil{Dipartimento di Fisica, Sezione di Astronomia, Universit\`a di Trieste, via G.B. Tiepolo 11, I-34131, Trieste, Italy} 
\author{K. Cunha}
\affil{Observat\'{o}rio Nacional, Rua General Jos\'{e} Cristino, 77, 20921-400 S\~ao Crist\'{o}v\~ao, Rio de Janeiro, RJ, Brazil}
\author{V. Smith and K. Hinkle}
\affil{National Optical Astronomy Observatory, 950 North Cherry Avenue, Tucson, AZ 85719, USA}
\author{M. Schultheis}
\affil{Observatoire de la Cote d'Azur, Boulevard de l'Observatoire, B.P. 4229, F 06304 NICE Cedex 4, France}

\altaffiltext{1}{Visiting astronomer, Kitt Peak National Observatory, National Optical Astronomy Observatory, which is operated by the Association of Universities for Research in Astronomy (AURA) under a cooperative agreement with the National Science Foundation.}

\begin{abstract}
Asymptotic Giant Branch stars are known to produce `cosmic' fluorine but it is uncertain whether these stars are the main producers of fluorine in the Solar neighborhood or if any of the other proposed formation sites, type II supernovae and/or Wolf-Rayet stars, are more important.  Recent articles have proposed both Asymptotic Giant Branch stars as well as type II supernovae as the dominant sources of fluorine in the Solar neighborhood.

In this paper we set out to determine the fluorine abundance in a sample of 49 nearby, bright K-giants for which we previously have determined the stellar parameters as well as alpha abundances homogeneously from optical high-resolution spectra. The fluorine abundance is determined from a 2.3~$\mu$m HF molecular line observed with the spectrometer Phoenix.

We compare the fluorine abundances with those of alpha elements mainly produced in type II supernovae and find that fluorine and the alpha-elements do not evolve in lock-step, ruling out type II supernovae as the dominating producers of fluorine in the Solar neighborhood.

Furthermore, we find a secondary behavior of fluorine with respect to oxygen, which is another evidence against the type II supernovae playing a large role in the production of fluorine in the Solar neighborhood. This secondary behavior of fluorine will put new constraints on stellar models of the other two suggested production sites: Asymptotic Giant Branch stars and Wolf-Rayet stars.
\end{abstract}


\section{Introduction} \label{sec:introduction}

Fluorine is an element with an intriguing and uncertain cosmic
origin. It has only one stable isotope, $^{19}$F, and
is easily destroyed in stellar interiors demanding intricate chains
of events for any created fluorine to survive from one stellar
generation to the next. The fluorine production in `normal'
explosive nucleosynthesis in Type Ia and II SNe is therefore believed
to be small. Three possible formation sites that will provide
possible fluorine production chains have been proposed:

\begin{itemize}
\item Asymptotic Giant Branch (AGB) are suggested to produce fluorine by means of two different processes. At solar metallicity the main chain of F production starts with $^{14}$N and includes $\alpha$, neutron, and proton captures. However, at slightly lower metallicities, the fluorine production is suggested to be dependent on $^{13}$C \citep{2014A&A...570A..46C}.

\item Type II Supernovae (SNeII) are suggested to produce cosmic fluorine via the neutrino process. The enormous quantity of neutrinos released during core-collapse can possibly, in spite of the small neutrino cross-section, turn a significant amount of the $^{20}$Ne in the outer envelopes of the collapsing star into fluorine \citep{1988Natur.334...45W}.

\item Wolf-Rayet (WR) stars might deposit fluorine into the interstellar medium via their strong stellar winds \citep{2000A&A...355..176M,2005A&A...443..243P}. Just like in the AGB-scenario, the fluorine in WR winds is produced in reactions starting from $^{14}$N, including $\alpha$, neutron, and proton captures. Due to the strong metallicity-dependence on the winds of these stars, a possible fluorine production through this channel is expected to start first at slightly sub-solar metallicities and then increase for higher metallicities \citep{2004MNRAS.354..575R}.
\end{itemize}

In addition to these scenarios, \citet{2011ApJ...737L..34L} show that fluorine can be produced by white dwarfs merging forming hydrogen-deficient stars with high carbon and fluorine abundances. However, the low number of known stars of this kind compared to the AGB stars, SNeII, and WR stars, makes it probable that the contribution of `cosmic' fluorine by this process compared to the others listed above is small.
 
The production of fluorine by AGB stars has been observationally
proven
\citep{1992A&A...261..164J,2009ApJ...694..971A,2010ApJ...715L..94A,2005A&A...433..641W,2005ApJ...631L..61Z,2008ApJ...682L.105O,2007ApJ...667L..81S,2011ApJ...729...40L,2011A&A...536A..40A}
but it is not certain that the fluorine produced by this process
can account for the cosmic abundance. Recently we made first use
of a
HF molecular line at 12.2 $\mu$m to determine the
fluorine abundance in a handful stars in the Solar neighborhood
\citep{2014ApJ...789L..41J}. By comparing our fluorine and oxygen
abundances to the chemical evolution models of
\citet{2011ApJ...739L..57K}, we proposed that the production of
fluorine by AGB-stars might be enough to explain the fluorine
abundance in the Solar neighborhood. However, since then
\citet{2015AJ....150...66P} observed a much larger sample than
ours, 51 stars, and draw the conclusion, based on the same chemical
evolution models of \citet{2011ApJ...739L..57K}, that the $\nu$-process
likely is the dominant producer of fluorine in the Solar neighborhood.
Furthermore, \citet{2015A&A...581A..88A} present new chemical
evolution models where the amount of fluorine produced by AGB-stars
in the Solar neighborhood is predicted to be much lower than in
the models of \citet{2011ApJ...739L..57K}, thereby questioning
models used and conclusions drawn in our paper and in
\citet{2015AJ....150...66P}.

One way of determining the possible importance of the $\nu$-process
in fluorine production in the Solar neighborhood purely
experimentally would be to determine fluorine and oxygen abundances
in a stellar sample large enough to trace a reliable trend in the
F-O abundances.  Oxygen is produced mainly in SNeII and if the
$\nu$-process is the dominant producer of fluorine the evolution
of oxygen and fluorine abundances would follow each other.

Another way of empirically testing the importance of the different fluorine production channels would be to look for possible secondary behaviors in the fluorine abundance as compared to other abundances.
Regarding the suggested fluorine production chain for solar-metallicity AGB stars it starts with $^{14}$N, an element which in turn can be both primary and secondary. In fact, $^{14}$N can be produced in a primary way if the C and O progenitors are produced by the star itself and not present in the star since its birth (secondary case). This happens in AGB stars during the third dredge-up episode \citep[e.g.,][]{1981A&A....94..175R,1999A&A...344..123M,2010MNRAS.403.1413K}.
Regarding the suggested fluorine production in lower metallicity AGB stars, it depends on the amount of $^{13}$C available (which can also be either secondary or primary) in the helium-rich inter-shell  but also on the amount of $^{13}$C in the ashes of the hydrogen-burning shell. While the former is weakly dependent on metallicity, the latter scales with the CNO abundances in the envelope which in turn depend upon the $^{12}$C made in the helium-burning shell and dredge-up up during the third dredge-up. Because of this, larger [F/Fe] ratios are predicted by metal-poor low-mass AGB stellar models as compared to solar-metallicity models \citep{2011ApJS..197...17C}.
Regarding the fluorine production in WR stars, it is also suggested to have the same duality according  to the secondary or primary nature of $^{14}$N. In this case $^{14}$N is generally produced as a secondary element except in the case of rapid stellar rotation which acts in the same way as the the third dredge-up. However, in this case a substantial primary $^{14}$N production is restricted to very metal-poor stars (Z$<10^{-5}$) as is suggested by several papers of the Geneva group \citep[e.g.,][]{2002A&A...381L..25M,2003A&A...411..543M,2003A&A...410..257C,2006A&A...449L..27C} and by \citet{2002ApJ...577..281C,2004ApJ...608..405C} in zero-metallicity star models. The necessity of some primary $^{14}$N from massive stars was claimed first by \citet{1986MNRAS.221..911M} to reproduce the [N/O] ratios observed in halo stars. To summarize, the uncertainties in the yields of massive stars for $^{14}$N and fluorine are still large and the yields of rotating massive stars are restricted to extremely metal poor stars.

To conclude, the primary and/or secondary nature of $^{14}$N in the production chains above, will in turn, if any of these production channels is the dominant one, make fluorine showing an abundance primary and/or secondary to oxygen: more oxygen results in more $^{14}$N and consequently more fluorine will be produced.

When it comes to the $\nu$-process, the fluorine possibly produced via this channel would be of primary nature, since it will depend on $^{20}$Ne that is ultimately originating from helium, whose abundance in turn is not dependent on the metallicity of the star.  

An experimental test of the possible secondary nature of fluorine with respect to oxygen will therefore help to constrain stellar models regarding the effectivity of different mixing scenarios in the AGB and WR stars. Furthermore, a possible secondary behavior of fluorine with respect to oxygen would not be present if the $\nu$-process is the main producer of fluorine in the Solar neighborhood, in that case ruling this production channel out completely.

In this paper we conduct the two experimental tests of the $\nu$-process described above by looking for a possible flat trend in the [F/O] vs. [O/H]-plane and investigating the possible primary and/or secondary nature of fluorine compared to oxygen. 

\section{Observations} \label{sec:observations}

We selected 49 K-giants from the Solar neighborhood sample of giants
in \citet{paperi}, with temperatures low enough to show a HF-line
at 2.3~$\mu$m.  The basic data of our stars are listed in \citet{paperi}.
In \citet{paperi} the stars were analyzed based on optical spectra
taken with the spectrometer FIES \citep{2014AN....335...41T}
($R\sim65000$) at the Nordic Optical Telescope.  Stellar parameters
as well as oxygen, magnesium, calcium, and titanium abundances were
determined from these spectra.

The stars chosen from this sample for the present project were
observed using the spectrometer Phoenix \citep{1998SPIE.3354..810H}
at the 4m Mayall telescope in two standard settings, one covering
the HF-line at 2.34~$\mu$m and one covering the OH-lines around
1.56~$\mu$m. The Phoenix data were reduced following standard,
recommended, procedures\footnote{https://www.noao.edu/kpno/phoenix/}
including removal of the telluric lines in the 2.3~$\mu$m-setting.

\section{Analysis} \label{sec:analysis}

The stellar parameters, as well as oxygen abundances from the [O
I]-line at 6300 \AA, were determined from optical FIES-spectra and
are taken from \citet{paperi}. This analysis is described in depth in
\citet{paperi}. In short, the stellar parameters are derived
purely spectroscopically from fitting weak Fe I and Fe II lines and
the wings of strong Ca I lines. All spectra have high S/N (typically
around 100).  Care was taken to calibrate
the stellar parameters against benchmark stars with well-known
stellar parameters (effective temperatures from angular diameter
measurements and surface gravities from asteroseismology).

Regarding the IR Phoenix-spectra the spectral line data and analysis
are the same as in \citet{2014A&A...564A.122J} for both the
HF-line at 2.34~$\mu$m and the OH-lines around 1.56~$\mu$m. 
As in that paper the program \texttt{Spectroscopy Made Easy}
\citep{1996A&AS..118..595V} was used together with a grid of
spherical symmetric, [$\alpha$/Fe]-enhanced, LTE MARCS-models to
determine the oxygen and fluorine abundances via $\chi^2$-fitting
of synthetic spectra with different abundances.

As can be seen in Table \ref{tab:uncertainties}, the derived fluorine abundance is very sensitive to the adopted effective temperature of the star, and this is the main source of uncertainty in our investigation.
\begin{deluxetable}{lrrr}
\tablecaption{Uncertainties of the derived abundances of a typical star (HIP68567) for changes in the derived stellar parameters.\label{tab:uncertainties}}
\tablehead{
\colhead{Uncertainty} & \colhead{$\delta$ A(O)$_{\mathrm{[O I]}}$} & \colhead{$\delta$A(O)$_{\mathrm{OH}}$}& \colhead{$\delta$ A(F)}
}
\startdata
	    $\delta T_{\mathrm{eff}}=+50$~K &    +0.01   &    +0.08  &    +0.12 \\
	    $\delta \log g=+0.15$           &    +0.06   &   $-0.01$ & $\pm0.00$\\
	    $\delta$[Fe/H]$=+0.05$          &    +0.02   &    +0.02  &    +0.03 \\
	    $\delta v_{\mathrm{mic}}=+0.1$  & $\pm0.00$  & $\pm0.00$ & $\pm0.00$\\
\enddata    
\end{deluxetable}

\section{Results} \label{sec:results}

The stellar parameters determined, oxygen abundances from the optical [O
I]-line at 6300 \AA~ as well as from the 1.56~$\mu$m OH-lines, and
fluorine abundances for our program stars are listed in  Table
\ref{tab:abunds}.

\begin{deluxetable}{ccccccccc}
\tablecaption{Stellar parameters and determined abundances for our program giants.\label{tab:abunds}}
\tablehead{
\colhead{Star} & \colhead{$T_{\mathrm{eff}}$} & \colhead{$\log g$} & \colhead{[Fe/H]\tablenotemark{a}} & \colhead{$v_{\mathrm{mic}}$} & \colhead{A(O)$_{\mathrm{[O I]}}$} & \colhead{A(O)$_{\mathrm{OH}}$} & \colhead{A(O)$^{\mathrm{b}} _{\mathrm{mean}}$} & \colhead{A(F)}
}
\startdata
    HIP48455 &   4461 &    2.65 &    0.20 &    1.55 &    8.93 &    8.91 &    8.92 &    4.76 \\
    HIP68501\tablenotemark{c} &   3925 &    1.32 &   -0.50 &    1.41 &    8.49 &    8.43 &    8.46 &    3.70  \\
    HIP68567 &   4190 &    1.95 &   -0.17 &    1.46 &    8.64 &    8.64 &    8.64 &    4.22 \\
    HIP69067 &   3928 &    1.27 &   -0.20 &    1.47 &    8.47 &    8.53 &    8.50 &    4.02 \\
    HIP69118 &   4152 &    1.86 &   -0.21 &    1.42 &    8.61 & \nodata &    8.61 &    4.23 \\
    HIP69316 &   4433 &    2.70 &    0.24 &    1.48 & \nodata &    8.96 &    8.96 &    4.81 \\
    HIP70899\tablenotemark{c} &   3921 &    1.48 &    0.05 &    1.37 & \nodata & \nodata & \nodata &    4.52 \\
    HIP70949 &   4085 &    1.60 &   -0.30 &    1.52 &    8.54 &    8.54 &    8.54 &    4.09 \\
    HIP72499\tablenotemark{c} &   4440 &    2.68 &    0.30 &    1.55 &    8.94 &    8.96 &    8.95 &    4.91 \\
    HIP73203\tablenotemark{c} &   4044 &    1.49 &   -0.58 &    1.43 &    8.52 &    8.42 &    8.47 &    3.63 \\
    HIP73568 &   3933 &    1.26 &   -0.14 &    1.48 &    8.51 &    8.55 &    8.53 &    4.15 \\
    HIP73917 &   4216 &    1.96 &   -0.14 &    1.52 &    8.60 & \nodata &    8.60 &    4.25 \\
    HIP75541 &   4123 &    1.84 &   -0.07 &    1.50 & \nodata & \nodata & \nodata &    4.20 \\
    HIP75572 &   4019 &    1.39 &   -0.45 &    1.48 &    8.42 &    8.42 &    8.42 &    3.82 \\
    HIP75583 &   4177 &    1.71 &   -0.42 &    1.50 &    8.54 & \nodata &    8.54 &    4.07 \\
    HIP76634 &   4100 &    2.13 &    0.16 &    1.42 & \nodata & \nodata & \nodata &    4.68 \\
    HIP77743 &   4404 &    2.48 &    0.24 &    1.42 &    8.85 & \nodata &    8.85 &    4.70 \\
    HIP78157 &   4439 &    2.67 &    0.24 &    1.58 &    8.93 & \nodata &    8.93 &    4.90 \\
    HIP78262 &   4049 &    1.77 &   -0.03 &    1.46 &    8.65 & \nodata &    8.65 &    4.37 \\
    HIP79120 &   4051 &    1.91 &    0.05 &    1.41 &    8.70 &    8.77 &    8.73 &    4.56 \\
    HIP79488 &   4030 &    1.54 &   -0.13 &    1.58 &    8.54 &    8.58 &    8.56 &    4.16 \\
    HIP79953 &   4126 &    1.68 &   -0.20 &    1.49 &    8.59 &    8.60 &    8.59 &    4.16 \\
    HIP80693 &   4096 &    1.95 &    0.07 &    1.61 &    8.77 &    8.75 &    8.76 &    4.50 \\
    HIP81119 &   3923 &    1.24 &   -0.26 &    1.48 & \nodata & \nodata & \nodata &    3.88 \\
    HIP82012 &   4061 &    1.55 &   -0.25 &    1.49 &    8.54 &    8.54 &    8.54 &    4.08 \\
    HIP82611 &   4193 &    1.70 &   -0.48 &    1.54 &    8.53 &    8.48 &    8.50 &    3.89 \\
    HIP82802\tablenotemark{c} &   4075 &    1.73 &   -0.16 &    1.45 &    8.61 &    8.64 &    8.62 &    4.25 \\
    HIP83677 &   4066 &    1.66 &   -0.12 &    1.52 &    8.60 &    8.62 &    8.61 &    4.12 \\
    HIP84431 &   4253 &    1.98 &   -0.06 &    1.58 &    8.68 &    8.68 &    8.68 &    4.41 \\
    HIP84659 &   4355 &    2.01 &   -0.19 &    1.55 & \nodata & \nodata & \nodata &    4.40 \\
    HIP84950 &   3954 &    1.11 &   -0.27 &    1.53 &    8.42 &    8.47 &    8.44 &    3.80 \\
    HIP85109 &   4274 &    2.31 &    0.05 &    1.35 & \nodata & \nodata & \nodata &    4.55 \\
    HIP85692 &   4107 &    1.59 &   -0.22 &    1.54 &    8.55 &    8.56 &    8.55 &    3.99 \\
    HIP85824\tablenotemark{c} &   4080 &    1.68 &   -0.36 &    1.37 & \nodata & \nodata & \nodata &    4.01 \\
    HIP85838 &   3936 &    1.61 &    0.09 &    1.57 &    8.64 & \nodata &    8.64 &    4.56 \\
    HIP86667\tablenotemark{c} &   3966 &    1.59 &   -0.19 &    1.38 &    8.59 &    8.56 &    8.57 &    4.13 \\
    HIP87445 &   4207 &    1.78 &   -0.26 &    1.53 &    8.58 &    8.58 &    8.58 &    4.16 \\
    HIP87777 &   4387 &    2.21 &   -0.02 &    1.49 & \nodata &    8.75 &    8.75 &    4.51 \\
    HIP88770 &   4066 &    1.64 &   -0.25 &    1.50 &    8.56 &    8.57 &    8.56 &    4.13 \\
    HIP88877 &   4012 &    1.46 &   -0.27 &    1.46 & \nodata &    8.50 &    8.50 &    4.04 \\
    HIP89298\tablenotemark{c} &   4012 &    1.34 &   -0.36 &    1.49 &    8.44 &    8.45 &    8.44 &    3.91 \\
    HIP89827 &   4286 &    2.06 &   -0.10 &    1.59 & \nodata &    8.69 &    8.69 &    4.55 \\
    HIP90915 &   4018 &    1.40 &   -0.18 &    1.61 &    8.54 &    8.58 &    8.56 &    3.95 \\
    HIP92768 &   4147 &    1.84 &   -0.19 &    1.42 & \nodata &    8.65 &    8.65 &    4.24 \\
    HIP93256 &   4331 &    2.09 &   -0.28 &    1.39 &    8.58 &    8.58 &    8.58 &    4.28 \\
    HIP93488 &   4129 &    1.61 &   -0.18 &    1.75 &    8.64 &    8.65 &    8.64 &    4.13 \\
    HIP94591 &   4216 &    1.83 &   -0.18 &    1.57 &    8.64 & \nodata &    8.64 &    4.19 \\
    HIP96063 &   4215 &    2.10 &   -0.07 &    1.46 & \nodata & \nodata & \nodata &    4.44 \\
    HIP97789 &   4110 &    1.73 &   -0.05 &    1.60 &    8.69 & \nodata &    8.69 &    4.24 \\
\enddata    
\tablenotetext{a}{We use A(Fe)$_{\odot}$=7.50 \citep{2009ARA&A..47..481A}.}
\tablenotetext{b}{The average dispersion between the oxygen abundance as derived from the [O I]- and OH-lines is 0.03 dex.}
\tablenotetext{c}{Star with probable ($>$30~\%) thick disk type kinematics and/or magnesium and titanium abundances of thick disk type.}
\end{deluxetable}

\begin{figure*}[!h]
\epsscale{2.0}
\plotone{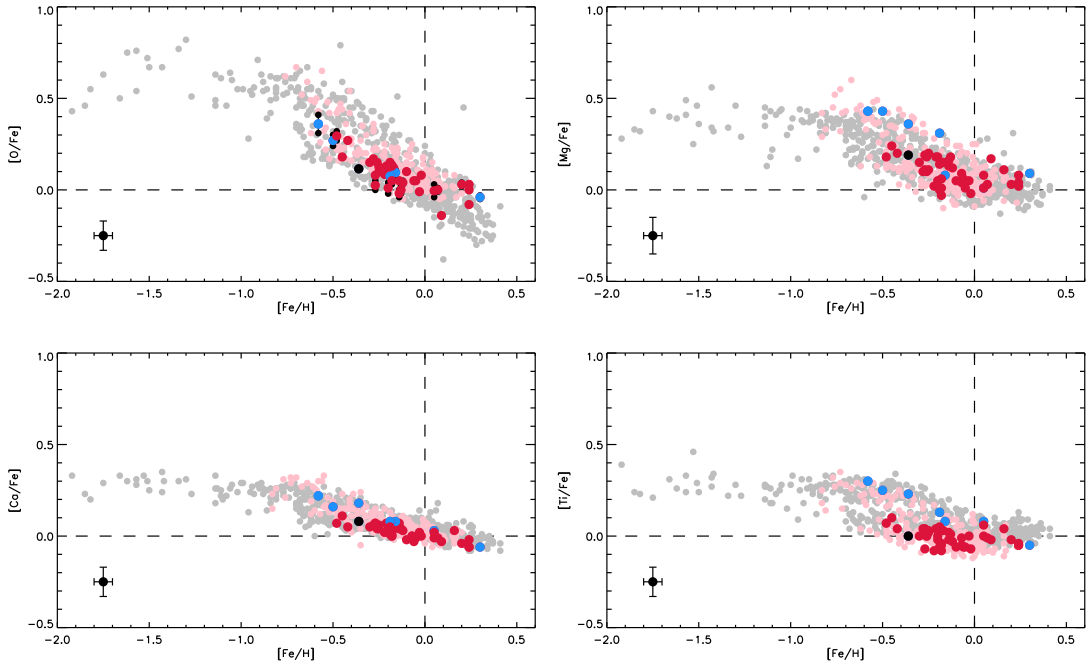}
\caption{[O/Fe], [Mg/Fe], [Ca/Fe], and [Ti/Fe] versus [Fe/H] taken from \citet{paperi} for our program stars in red and blue for probable thin and thick disk stars, respectively. The one star, HIP89298, that is a kinematically probable halo star is marked in black. For the stars where we have oxygen abundances derived from both the [O I] 6300~\AA~line and the 1.55~$\mu$m OH lines the two values are marked in black with the mean in red or blue. The pink dots are the abundances of the complete giant star sample of \citet{paperi} and the grey dots are abundances from solar neighborhood dwarf stars \citep{2014A&A...562A..71B}. We use A(O)$_{\odot}=8.69$, A(Mg)$_{\odot}=7.60$, A(Ca)$_{\odot}=6.34$, A(Ti)$_{\odot}=4.95$, and A(Fe)$_{\odot}=7.50$ \citep{2009ARA&A..47..481A}.\label{fig:alfa}}
\end{figure*}

\begin{figure}[!h]
\epsscale{1.1}
\plotone{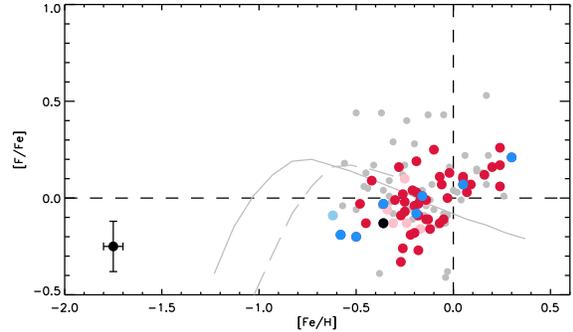}
\caption{[F/Fe] versus [Fe/H] for our program stars, red for probable thin disk stars and blue for probable thick disk stars. The one star, HIP89298, that is kinematically a probable halo star is marked in black. The fluorine abundances from \citet{2014ApJ...789L..41J} as derived from the 12.2~$\mu$m HF-line are shown in pink and lighter blue for thin and thick disk stars, respectively, and the values from \citet{2015AJ....150...66P} are plotted in grey. The grey line is the thin disk chemical evolution model of \citet{2011MNRAS.414.3231K} based on fluorine production in AGB-stars only, and the dashed line is the corresponding chemical evolution model for the thick disk. We use A(F)$_{\odot}=4.40$ \citep{2014ApJ...788..149M}, and A(Fe)$_{\odot}=7.50$ \citep{2009ARA&A..47..481A}.\label{fig:ffe}}
\end{figure}

\begin{figure*}[!h]
\epsscale{2.2}
\plotone{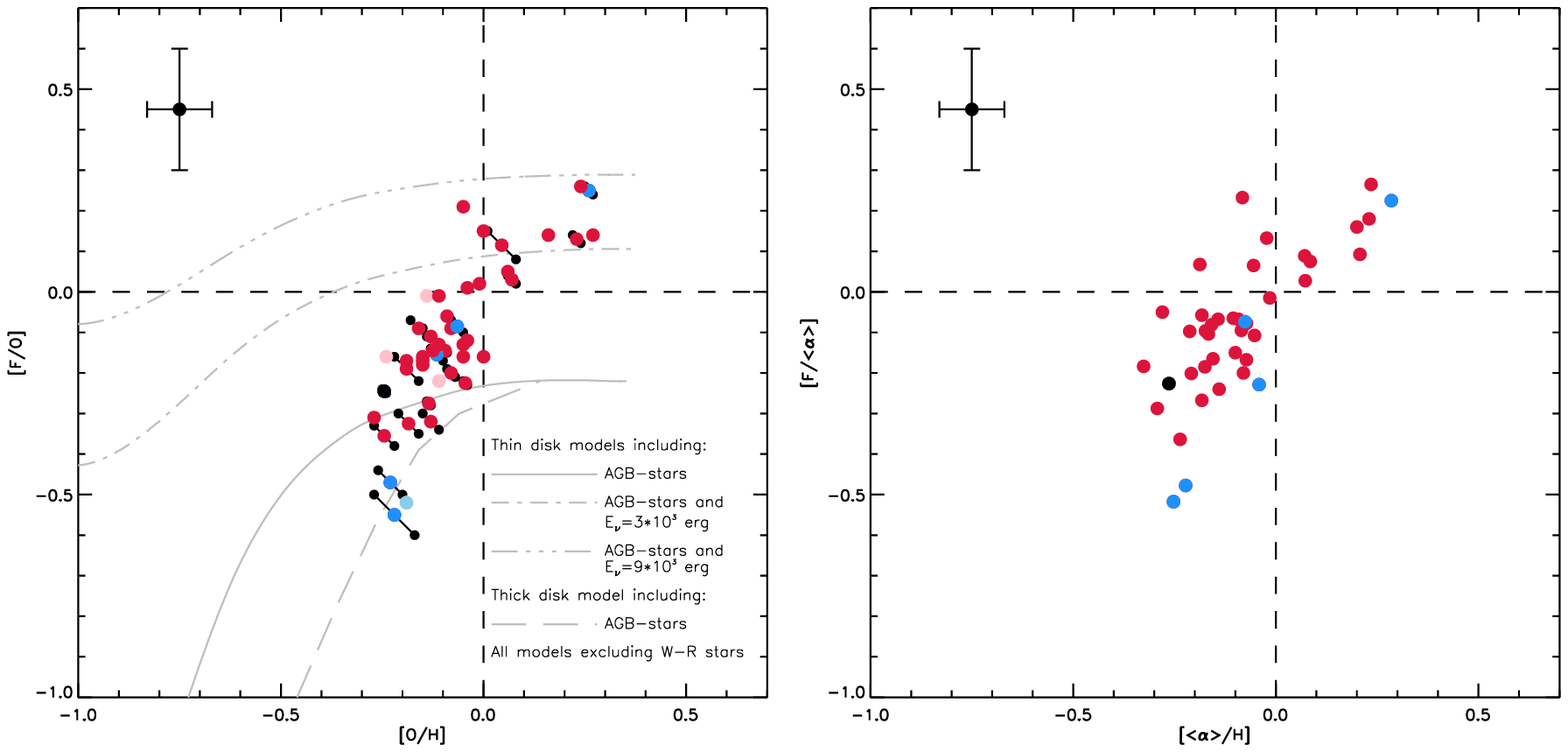}
\caption{[F/O] versus [O/H] and [F/$<\alpha >$] versus [$<\alpha >$/H] for our program stars, red for probable thin disk stars and blue for probable thick disk stars. The one star, HIP89298, that is kinematically a probable halo star is marked in black. The fluorine abundances from \citet{2014ApJ...789L..41J} as derived from the 12.2~$\mu$m HF-line are shown in pink and lighter blue for thin and thick disk stars, respectively. The grey lines show different chemical evolution models of \citet{2011ApJ...739L..57K,2011MNRAS.414.3231K}. We use A(O)$_{\odot}=8.69$ \citep{2009ARA&A..47..481A}, and A(F)$_{\odot}=4.40$ \citep{2014ApJ...788..149M}.\label{fig:fo}}
\end{figure*}

In Figure \ref{fig:alfa} we show the alpha elemental abundances for
our stars as determined in \citet{paperi}.  In Figure \ref{fig:ffe}
we show [F/Fe] versus [Fe/H].  In Figure \ref{fig:fo} we
show [F/O] versus [O/H] and [F/$<\alpha >$] versus [$<\alpha
>$/H].  In Figure \ref{fig:secondary} we show the A(F)
versus A(O) for the program stars.

Following the scheme in \citet{2015AJ....150...66P}, in turn from
\citet{2013ApJ...764...78R} and \citet{1987AJ.....93..864J}, we estimate the
kinematic probability that our stars belong to the thin disk, thick
disk, and/or halo respectively (see Table \ref{tab:abunds}). Stars
with more than 30~\% probability to kinematically belong to the
thick disk and/or have magnesium and titanium abundances of thick
disk type, are marked in blue in the figures. One star, HIP89298,
is kinematically a probable halo star (53~\%) and is marked in
black. Its alpha abundance, however, makes it most probable that
it belongs to the thin disk. The division into thin and thick
disk chemical composition is most clearly seen in magnesium and
titanium.  This is less obvious in calcium.  It is not seen at all
in oxygen for unknown reasons. As can be seen in the pink dots in
Figure \ref{fig:alfa}, there are many stars with typical thick-disk
type oxygen abundances in the \citet{paperi} sample but just not
in the stars analyzed here for fluorine.  The
fact that we derive very similar oxygen abundances from the 6300
\AA~[O I]-line and the 1.56 $\mu$m OH-lines makes us confident that
our oxygen abundances do not have the large systematic uncertainties
one might suspect from looking at Figure \ref{fig:alfa}.

\section{Discussion} \label{sec:discussion}

From Figures \ref{fig:ffe}-\ref{fig:fo} it is obvious that none of the chemical evolution models from \citet{2011ApJ...739L..57K,2011MNRAS.414.3231K} reproduce our observational trends. Similarly neither would the very similar models of \citet{2004MNRAS.354..575R} or the AGB-only models of \citet{2015A&A...581A..88A} that predict much lower [F/Fe] compared to [Fe/H] than the models shown in Figure \ref{fig:ffe}.

\begin{figure}[!h]
\epsscale{1.1}
\plotone{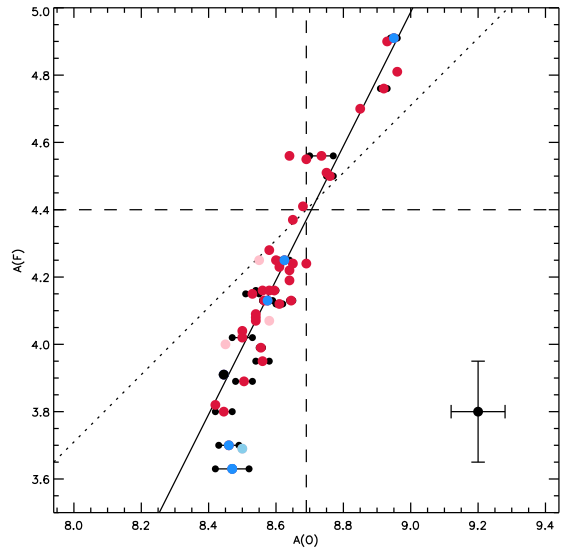}
\caption{A(F) versus A(O) for our program stars, red for probable thin disk stars and blue for probable thick disk stars. The one star, HIP89298, that is a kinematically probable halo star is marked in black. The fluorine abundances from \citet{2014ApJ...789L..41J} as derived from the 12.2~$\mu$m HF-line are shown in pink and lighter blue for thin and thick disk stars, respectively. The dashed lines mark the solar values used and the dotted line shows a solar-scaling between F and O (a slope of one). The solid line shows a linear fit to the thin disk data and has a slope of 2.0.\label{fig:secondary}}
\end{figure}

The fact that the locus of points in Figures \ref{fig:ffe}-\ref{fig:secondary} more or less passes through the solar values gives credibility to the accuracy of the solar fluorine abundance of A(F)$_{\odot}=4.40\pm0.25$ determined using the K-band spectrum of a sunspot \citep{2014ApJ...788..149M}.

Our [F/Fe] versus [Fe/H]-trend in Figure~\ref{fig:ffe} is more scattered than the alpha-trends of Figure \ref{fig:alfa}, which is expected from the uncertainties listed in Table \ref{tab:uncertainties}. Still, our data show a growing trend over the observed metallicities: there are no points in the fourth quadrant of the plot, all the five lowest metallicity stars have sub-solar [F/Fe], and a non-parametric test shows a linear relation is a best guess. A linear regression to our data yields a fit with a positive inclination of 0.4, with a Pearson’s correlation of +0.6.
This means that [F/Fe] is increasing in the same [Fe/H]-range where the [$\alpha$/Fe] are seen to diminish due to the onset of SNeIa producing large amounts of iron. The source producing fluorine at these times must therefore produce relatively more fluorine than the iron produced in SNeIa. If the $\nu$-process was the dominant fluorine producer, one would on the contrary expect fluorine to qualitatively show an alpha-like trend with respect to iron, because both fluorine and the alpha elements would in that case be mainly  produced in SNeII (and not in SNeIa). On the other hand the rising [F/Fe] versus [Fe/H]-trend qualitatively supports the creation of fluorine in AGB stars and/or WR stars. The contribution of fluorine from the low-mass, fluorine-producing,  AGB-stars is expected for rather high [Fe/H] because of their long life times. 
The formation rate for WR-stars is higher for higher metallicity and the mass-loss rate of WR-stars increases with metallicity. Therefore, fluorine production by WR-stars would be expected to be very metallicity dependent. However, the amount of `cosmic' fluorine produced by these stars is also dependent on the rotation of the star, and it is not known if the rotational velocity distribution of WR-stars is dependent on metallicity, making the fluorine yield by this source very uncertain.

In the models of \citet{2011MNRAS.414.3231K}, the thin and thick disk are predicted to separate with the thick disk having lower [F/Fe] for the lowest metallicities. This is not seen in our data, if anything, the probable thick disk stars are on the contrary showing higher [F/Fe] than the thin disk stars for the lowest metallicities in our sample.

Our observational trend corroborates results from \citet{2014ApJ...789L..41J} but the larger size of the present sample allows us to reveal the surprising trend of growing [F/Fe]. Our results also roughly overlap those of  \citet{2015AJ....150...66P} in the [F/Fe] versus [Fe/H] plane but with a smaller scatter leading us to a different conclusion. The smaller scatter in our data is expected since \citet{2015AJ....150...66P} use stellar parameters from different literature sources while we derive them homogeneously using optical spectra \citep{paperi}.

In the left panel of Figure \ref{fig:fo} we have plotted [F/O]
versus [O/H] to better see the possible influence of the $\nu$-process.
If the $\nu$-process was the dominating producer of fluorine in the
solar neighborhood, [F/O] is expected to be more or less constant
for all [O/H]. This is definitively not the case, with our observational
trend showing a similar steep increase of [F/O] with [O/H] as in
the AGB-only chemical evolution models of \citet{2011ApJ...739L..57K}.
However, the observational trend is much higher in [F/O] than what
is predicted by the models. This {\it might} hint at WR-stars having
a significant importance in the fluorine production in the Solar
neighborhood (since they are not included in the models of
\citet{2011ApJ...739L..57K}) but to fully understand and distinguish
between the different scenarios more modeling is needed. To rule
out the possibility that the trend is influenced by the non-thick-disk
behavior of our oxygen abundances we have replaced the oxygen
abundance with the mean alpha abundances in the right panel of
Figure \ref{fig:fo}. The conclusion is again the same. The evolution
of fluorine is not following the evolution of the alpha elements,
which are produced in SNeII, indicating that the $\nu$-process
cannot be the dominant producer of fluorine in the Solar neighborhood.

In Figure \ref{fig:secondary} we have plotted A(F) versus A(O)
to look for a possible secondary or primary behavior of fluorine
with respect to oxygen. Indeed, the linear fit to the thin disk
data shows a slope of 2.0 indicating a secondary behavior of
fluorine with respect to oxygen. If the $\nu$-process was the
dominant producer of fluorine in the thin disk, fluorine would
be primary with respect to oxygen and the slope would be one. This
secondary nature can only be explained if the fluorine produced
is dependent on the amount of oxygen available. This in turn
can be explained by the proposed fluorine production-channels in
AGB-stars and WR-stars starting with $^{14}$N that is itself secondary
with respect to oxygen \citep{2016MNRAS.458.3466V}, resulting in 
the secondary behavior of fluorine.

\section{Conclusions} \label{sec:conclusion}
We have determined the fluorine abundance in a sample of 49 nearby, bright K-giants for which we previously have determined the stellar parameters as well as alpha abundances homogeneously from optical high-resolution spectra \citep{paperi}.
Our observational [F/Fe] vs. [Fe/H] and [F/O] vs. [O/H] trends are both increasing, which they would not be if the $\nu$-process was the dominant fluorine producer in the Solar neighborhood, severely limiting its possible contribution to the cosmic fluorine budget.
We also find an empirical secondary relation between fluorine and oxygen further strengthening this case. This secondary nature will put new constraints on stellar models of AGB and WR stars.

\acknowledgments
This research is based on observations made at Kitt Peak National Observatory, National Optical Astronomy Observatory, which is operated by the Association of Universities for Research in Astronomy (AURA) under a cooperative agreement with the National Science Foundation, and on observations made with the Nordic Optical Telescope, operated by the Nordic Optical Telescope Scientific Association at the Observatorio del Roque de los Muchachos, La Palma, Spain, of the Instituto de Astrofisica de Canarias.
   
 This research has been partly supported by the Lars Hierta Memorial Foundation, Per Westlings minnesfond, and the Royal Physiographic Society in Lund through Stiftelsen Walter Gyllenbergs fond, and the travel grants for young researchers.
 
 This publication made use of the SIMBAD database, operated at CDS, Strasbourg, France, NASA's Astrophysics Data System, and the VALD database, operated at Uppsala University, the Institute of Astronomy RAS in Moscow, and the University of Vienna.

{\it Facilities:} \facility{Mayall (Phoenix)}, \facility{NOT (FIES)}.

\end{document}